\documentclass{aa}  

\usepackage{graphicx}
\usepackage{txfonts}
\usepackage{lipsum}
\usepackage[normalem]{ulem}
\usepackage{subcaption}         
\usepackage{lscape}             
\usepackage{placeins}           
\usepackage{xcolor}
\usepackage[linesnumbered,ruled,vlined]{algorithm2e}

\SetCommentSty{mycommfont}
\SetKwRepeat{Do}{do}{while}%
\SetKwFor{Rtreefor}{rtree\char`_for}{do}{endfor}

\newcommand{\yr}{{\mathrm{yr}}}
\newcommand{\au}{{\mathrm{au}}}
\newcommand{\Msun}{\ensuremath{\,\rm{M}_{\odot}}\xspace}
\newcommand{\Rsun}{\ensuremath{\,\rm{R}_{\odot}}\xspace}

\defcitealias{Marcello2017}{M17}
\defcitealias{Dehnen2000}{D00}
\defcitealias{Dehnen2002}{D02}
\defcitealias{Dehnen2014}{D14}
\usepackage{hyperref}
\usepackage[capitalise]{cleveref}
\usepackage{bm}

\begin{document}


   \title{Momentum-conserving self-gravity in {\sc Phantom}}


   \subtitle{Parallel dual tree traversal for the fast multipole method}


   \author{Yann Bernard\inst{1}
        \and Timothée David-Cléris\inst{1} \and Daniel J. Price\inst{2,1} \and Mike Y. M. Lau\inst{3,4}
        }

   \institute{Univ. Grenoble Alpes, CNRS, IPAG, 38000 Grenoble, France
         \and School of Physics and Astronomy, Monash University, VIC 3800, Australia
         \and
         Heidelberger Institut f\"{u}r Theoretische Studien, Schloss-Wolfsbrunnenweg 35, 69118 Heidelberg, Germany
        \and Zentrum f\"ur Astronomie der Universit\"at Heidelberg, Astronomisches Rechen-Institut, M\"onchhofstr. 12-14, 69120 Heidelberg 
             }

   \date{Received September 30, 20XX}

 
  \abstract
   {Tree codes that approximate groups of distant particles with multipole expansions accelerate gravitational force calculations. While momentum-conserving fast multipole methods exist, parallelisation is non-trivial inside a modern SPH framework.}
   {We aim for a parallel momentum and angular momentum-conserving Cartesian multipole method for the computation of the gravitational force in smoothed particle hydrodynamics (SPH) with adaptive gravitational force softening in {\sc Phantom}.} 
   {We modified Dehnen's original dual tree traversal algorithm to fit the parallel architecture of {\sc Phantom} by replicating the node-node interaction on the ancestors of each leaf node in the tree. While this notionally duplicates work, it greatly simplifies the parallelisation of the algorithm. We then recover the $\mathcal{O}(N)$ scaling of the algorithm by use of caches. We also adapt the tree opening criterion for adaptive softening lengths, such that all interactions within the softening kernel are handled pairwise (as in SPH) rather than with multipole expansions, also allowing the gravity calculation to be performed alongside the SPH force evaluation.}
   {We demonstrate that the new code conserves linear and angular momentum to machine precision while giving similar force accuracy and better computational performance to the previous (non-symmetric) self-gravity solver in {\sc Phantom}.}
   {Our new fast multipole method is now the default for computing self-gravity in the public code.}

   \keywords{gravitation ---
                hydrodynamics ---
                methods: numerical
               }

   \maketitle
   \nolinenumbers

\section{Introduction}
In the first applications of Smoothed Particle Hydrodynamics (SPH), \citet{Lucy1977} and \citet{Gingold1977} computed the self-gravity between binary stars modelled as gaseous polytropic spheres by direct $\mathcal{O}(N^2)$ summation. \citet{Monaghan1985} first accelerated this by interpolating to a grid and using a multigrid algorithm. 

\citet{Hernquist1989} and \citet{Benz1990} unified SPH with hierarchical tree codes, the former authors basing their method on the \citet{Barnes1986} Oct-tree, where the complexity is reduced to $\mathcal{O}(N\log N)$ by grouping distant particle-particle interactions into fewer `particle-node' interactions based on a critical opening angle between the particle and the size of the distant node. \citet{Benz1990} used a similar algorithm, but based on a nearest-neighbour binary tree rather than an Oct-tree.

These, and similar tree algorithms, with minor variation \citep[e.g.][]{Salmon1994}, have used in collisionless $N$-body and self-gravitating SPH codes ever since \citep[e.g.][]{Steinmetz1993,Dubinski1996,Springel2001,Wadsley2004,Springel2005,Wetzstein2009,Hubber2011,Gafton2011,Iwasawa2017,Wadsley2017,Price2018,Schaller2024}.

\citet{Greengard1987} first showed that $\mathcal{O}(N)$ scaling is possible using a Fast Multipole Method (FMM) where particle-node interactions are replaced by symmetric node-node interactions based on a spherical harmonic expansion. However, their algorithm was not widely adopted in astrophysics because it was found to be slower in practice (e.g.~\citealt{Capuzzo1998} found it to be almost 3 times slower than their tree code; see \citealt{Dehnen2014}).

A disadvantage of standard tree codes is that forces between particles are not equal and opposite, meaning that linear momentum is only approximately conserved. \citet{Dehnen2000,Dehnen2002} modified the fast multipole method to use symmetric Cartesian multipole expansions, explicitly conserving linear momentum to machine precision. But it requires a dual tree traversal algorithm to compute the gravitational force, which is recursive and has only been implemented to date with a task-based (or work-stealing) parallel paradigm \citep{Dehnen2014,Marcello_2021,Springel2021,Schaller2024}. The latter can be particularly complicated to implement in already optimised SPH codes. Achieving this kind of parallelism on GPUs is even more difficult, although this new platform is a clear goal for the future of astrophysics codes. \citet{Potter_2017} deliberately split the work between CPU and GPU to overcome this issue. However, in a fully GPU-based code like {\sc Shamrock} \citep{David--cleris_2025} this would be inefficient. Therefore another approach is needed. The FMM also need special treatment to use adaptive gravitational force softening to ensure that the resolution scales in the gas and gravitational force calculations are identical \citep[see e.g.][]{Bate1997,Price2007}.

In this paper, we describe a different parallel implementation of Dehnen's momentum-conserving tree code \citep{Dehnen2000} for the calculation of self-gravity in the {\sc Phantom} SPH code. The algorithm is designed for use with adaptive gravitational force softening and to be computed during the SPH force loop. It is faster than the previous self-gravity in {\sc Phantom} and comparable in accuracy. The paper is organised as follows: \Cref{sec:multipolemethod} covers mathematical background. \Cref{sec:num_impl} presents our new implementation. \Cref{sec:results} benchmarks it. We discuss in \Cref{sec:discuss} and conclude in \cref{sec:conclusion}.

%

\section{Review of multipole methods for self-gravity} \label{sec:multipolemethod}

For a collection of $N$ collisional point mass particles, the gravitational acceleration on a given particle $i$ is given by
\begin{equation}
\bm{a}_{i} =-G\sum_{j\neq i}^N \frac{ m_j}{\vert \bm{x}_i - \bm{x}_j \vert^3} (\bm{x}_i - \bm{x}_j), \label{eq:directsum}
\end{equation}
where $\bm{x}_i$ and $\bm{x}_j$ are the position vectors of the target and source particle, and $m_j$ is the mass of the source particle. 

For collisionless fluids or self-gravitating gas, the medium is a continuous density field where one solves for the gravitational potential from Poisson's equation
\begin{equation}
\nabla^2 \Phi = 4 \pi G \rho (\bm{x}).
\end{equation}
If the continuous density field $\rho (\bm{x})$ is represented by a set of particles of fixed mass, Poisson's equation has the solution
\begin{equation}
\Phi_{i} = -G \sum_{j=1}^N  m_j \phi(\vert \bm{x}_i - \bm{x}_j \vert, \epsilon), \label{eq:softdirectsum}
\end{equation}
where $\phi$ is the softening kernel, defined such that $\phi \to 1/r$ outside the compact support radius (typically several $r/\epsilon$). See \citet{Dehnen2001} and \cite{Price2007} for details.

\begin{figure}
    \centering
    \includegraphics[width=0.78\linewidth]{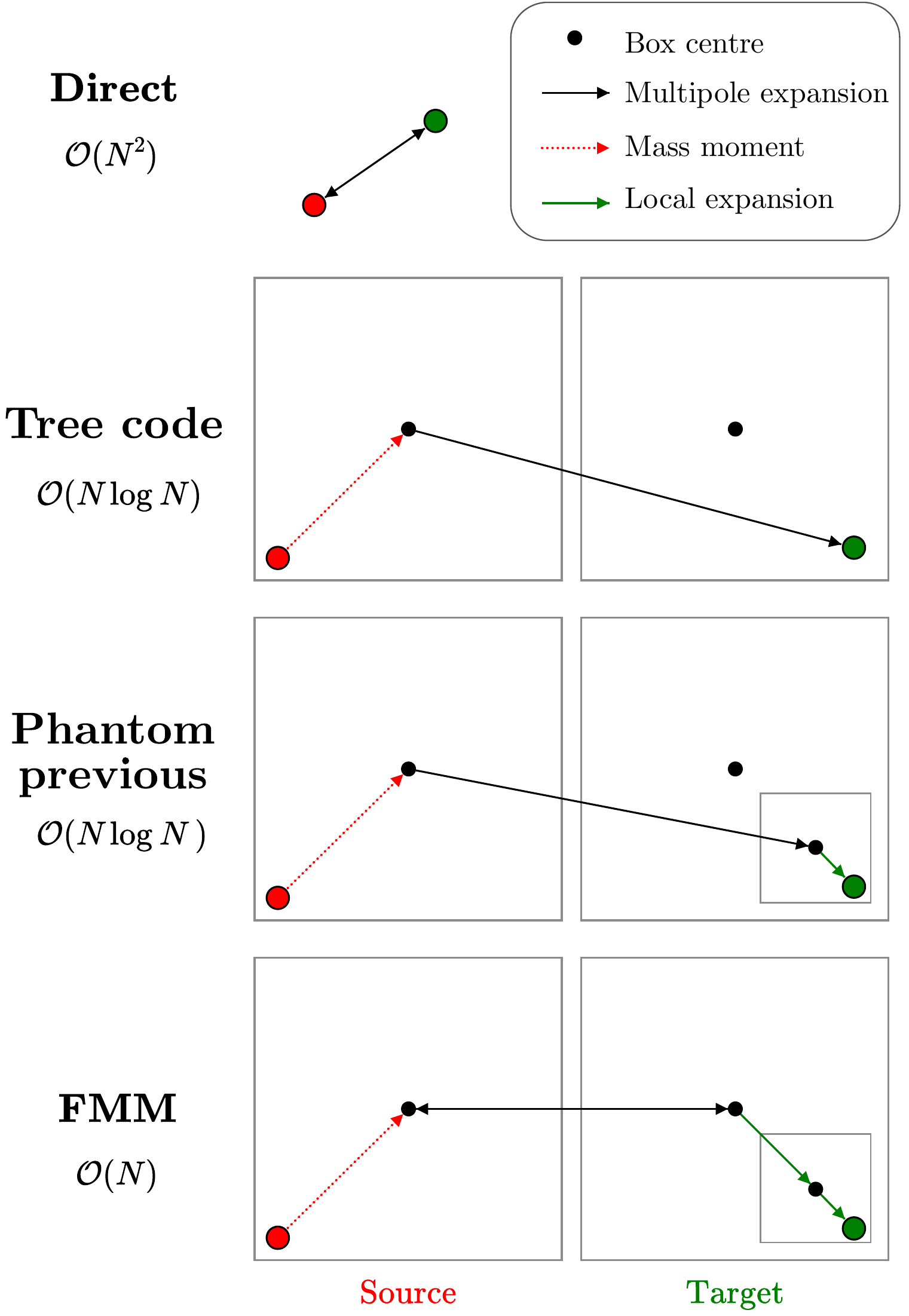}
    \caption{Multipole methods for self-gravity. Each row shows how a source particle (red, left) acts on a target particle (green, right). A direct sum (top row) computes all pair interactions. Tree codes (second row) evaluate the multipole expansion between the moments of the source node and the target particle. {\sc Phantom}'s previous scheme (third row) evaluated it between the source node and the target leaf node, with moments then translated to the particle. In FMM (bottom), the multipole expansion is evaluated once between node centres and applied in both directions, so that node--node interactions satisfy Newton's third law; force moments are then propagated down to particles.}
    \label{fig:summarymmmethods}
\end{figure}

In both \cref{eq:directsum,eq:softdirectsum} the force between two particles is equal and opposite, such that the total momentum $\sum_i m_i \bm{v}_i$ is conserved. But computing the acceleration of all $N_p$ particles via \cref{eq:directsum} requires $O(N_p^2)$ pairwise force evaluations (top row of \cref{fig:summarymmmethods}). In collisionless N-body and SPH simulations, where $N_p \gtrsim 10^6$, using \cref{eq:directsum} requires trillions of operations. Multipole methods aim to improve on $\mathcal{O}(N^2)$ performance while trying to conserve important quantities. 

\subsection{Multipole expansion of the gravitational acceleration}

In a self-gravitating medium, the solution of Poisson's equation at the position $\bm{x}_i$ can be expressed via the Green's function $g$ as
\begin{equation}
    \Phi (\bm{x_i}) = - G \int \rho(\bm{x}_j) g(\bm{x}_j - \bm{x}_i) {\rm d}^3 \bm{x}_j, \label{eq:poisson.integ}
\end{equation}
where $g(\bm{r}) = 1/\vert r\vert$, making the integral convergent only when convolved with the density field as per \cref{eq:softdirectsum}.

Introducing a third position $\bm{s}_b$ (to later represent the centre of a node), we can decompose $\bm{x}_j - \bm{x}_i$ into two new vectors, $\bm{r} \equiv \bm{s}_b - \bm{x}_i$ and $\bm{b}_j \equiv \bm{x}_j - \bm{s}_b$, such that $\bm{r}$ is the distance between the particle receiving the force and the node centre and $\bm{b}$ is that between a remote particle and the node centre.  Assuming  $\vert \bm{r} \vert \gg \vert \bm{b}_j \vert$, the Green's function can be approximated by its multipole expansion around $\bm{b}_j$, giving
\begin{equation}
    g(\bm{x}_j - \bm{x}_i) = g(\bm{r} + \bm{b}_j) \simeq \sum_{n = 0}^p \frac{1}{n!} \nabla_r^{(n)} g(\bm{r}) \cdot \bm{b}_j^{(n)},
\end{equation}
where $\nabla_r^{(n)}$ denotes the $n^{\rm th}$ gradient with respect to $r$ and $\bm{b}^{(n)}$ is the n-fold tensor product of $\bm{b}$. Here we truncated the expansion to order $p$ in powers of $\vert b_j  \vert /  \vert r \vert$. Using this in \cref{eq:poisson.integ} gives
\begin{align}
    \Phi(\bm{x}_i) &= - G\int_V \rho(\bm{x}_j) g (\bm{x}_j - \bm{x}_i) ~{\rm d}^3\bm{x}_j,\\
    &\simeq - G\int_V \rho(\bm{x}_j)  \sum_{n = 0}^p \frac{1}{n!} \nabla_r^{(n)} g(\bm{r}) \cdot \bm{b}_j^{(n)} ~{\rm d}^3\bm{x}_j,\\
    &= - G\sum_{n = 0}^p \frac{1}{n!} \underbrace{\nabla_r^{(n)} g(\bm{r})}_{\bm{D}_n} \cdot \underbrace{\left(\int_V \rho(\bm{x}_j) \bm{b}_j^{(n)}~{\rm d}^3\bm{x}_j\right)}_{\bm{Q}_n^B},
\end{align}
where $\bm{D}_n$ denotes the $n$th derivative of the Green's function and $\bm{Q}^B_n$ are the moments of the mass distribution.
The gravitational acceleration $\bm{a}_{\rm g}(\bm{x}_i)= -(\nabla_i \phi)(\bm{x}_i)$ can then be written as
\begin{align}
    \bm{a}_{\rm g}(\bm{x}_i) &= \nabla_i \int G \rho(\bm{x}_j)  g(\bm{x}_j - \bm{x}_i) {\rm d}^3\bm{x}_j, \\
    &= -G \int \rho(\bm{x}_j) \nabla_j g(\bm{x}_j - \bm{x}_i) {\rm d}^3\bm{x}_j, \\
    &= -G \sum_{n = 0}^p \frac{1}{n!} \bm{D}_{n+1} \cdot \bm{Q}_n^B.
\end{align}
In a system of massive particles, the $n$-th order mass moment created by a set of point-masses in a cartesian volume reduces to 
\begin{align}\label{eq:massmom}
  \bm{Q}_n^B = \sum_j m_j \bm{b}_j^{(n)}.
\end{align}

This expansion means that one can compute a single force vector on the particle $i$ using the multipole of a box containing multiple particles (see second row of \cref{fig:summarymmmethods}). Depending on the size of the box, this saves a large number of pairwise particle interactions. 
This expansion method combined with a hierarchical tree built on the particles is the most commonly used method in astrophysics as it is typically several orders of magnitude faster than a direct sum. This method is referred to as a ``tree code'', following \citet{Barnes1986}. In hydrodynamical simulations with millions or billions of particles, computing the multipole expansion on each individual particle is expensive.

\subsection{The Fast Multipole Method}
The Fast Multiple Method (FMM), introduced by \cite{Greengard1987} and reviewed by \cite{Dehnen2014}, reduces the number of evaluations of the Green's function to consider only node-node instead of particle-node interactions.

Consider two nodes at each side ($i$ and $j$) of the force evaluation. The distance vector between the particles $i$ and $j$, $\bm{x}_j - \bm{x}_i$ can be decomposed into three by introducing two new points $\bm{s}_a$ and $\bm{s}_b$, referring to the centres of the receiving and sending nodes. Keeping $\bm{b}_j$ as previously, we additionally define $\bm{a}_i \equiv \bm{x}_i - \bm{s}_a$ and $\bm{r} \equiv \bm{s}_b - \bm{s}_a$, such that
\begin{equation}
\bm{x}_j - \bm{x}_i = \bm{r} + (\bm{b}_j - \bm{a}_i).
\end{equation}
This reduces to the previous when taking a=0.
Under the assumption $\vert (\bm{b}_j - \bm{a}_i) \vert \ll \vert \bm{r} \vert$, we can perform an expansion with respect to $(\bm{b}_j - \bm{a}_i)$ instead of only $\bm{b}_j$ as done for the standard multipole expansion. Firstly, for two vectors $\bm{c}$, $\bm{d}$ multiplying a symmetric tensor $T$ of order larger than $n$, we have the identity
\begin{align}
     {\bf T} \cdot \left(\bm{d} + \bm{c}\right)^{(n)} &= {\bf T } \cdot \sum_{k=0}^n \binom{n}{k} \bm{d}^{(n-k)} {\bm c}^{(k)}.
\end{align}
Using this, the Green's function can be expanded according to
\begin{align}
    g(\bm{x}_j - \bm{x}_i) &= g(\bm{r} + \bm{b}_j - \bm{a}_i), \\
    &\simeq \sum_{n = 0}^p \frac{1}{n!} \nabla_r^{(n)} g(\bm{r}) \cdot \left(\bm{b}_j - \bm{a}_i\right)^{(n)}, \\
    &= \sum_{k = 0}^p \frac{(-1)^k}{k!} \bm{a}_i^{(k)} \cdot \sum_{n=0}^{p-k} \frac{1}{n!} \bm{D}_{n+k} \cdot \bm{b}_j^{(n)}.
\end{align}
Applying the same procedure as previously, we can insert this Green's function into \cref{eq:poisson.integ}, obtaining
\begin{align}
    \Phi(\bm{x}_i) &= -G \int_V \rho(\bm{x}_j) g (\bm{x}_j - \bm{x}_i) ~{\rm d}^3\bm{x}_j,\\
    &\simeq -G\int_V \rho(\bm{x}_j) \sum_{k = 0}^p \frac{(-1)^k}{k!} \bm{a}_i^{(k)} \cdot \sum_{n=0}^{p-k} \frac{1}{n!} \bm{D}_{n+k} \cdot \bm{b}_j^{(n)} ~{\rm d}^3\bm{x}_j, \\
    &= -G \sum_{k = 0}^p  \frac{\bm{a}_i^{(k)}}{k!} \cdot \underbrace{(-1)^k\sum_{n=0}^{p-k} \frac{1}{n!} \bm{D}_{n+k} \cdot {\bm{Q}_n^B}}_{\bm{M}_k},
\end{align}
where $\bm{M}_k$ is a rank-k tensor.
The corresponding acceleration is
\begin{align}
    \bm{a}_{\rm g}(\bm{x}_i) &= -G \sum_{k = 0}^p  \frac{\bm{a}_i^{(k)}}{k!} \cdot \underbrace{(-1)^k\sum_{n=0}^{p-k} \frac{1}{n!} \bm{D}_{n+k+1} \cdot {\bm{Q}_n^B}}_{{\rm d}\bm{M}_k}, \label{eq:dmkdeffmm}
\end{align}
where ${\rm d}\bm{M}_k$ is the derivative of $\bm{M}_k$ with respect to $r$.

\Cref{eq:dmkdeffmm} implies that the Green's function only needs to be evaluated once per pair of nodes. However, depending on the node associations used in the target selection algorithm, different flavours can be built using the same equation (\cref{eq:dmkdeffmm}). In the previous self-gravity implementation in {\sc Phantom} \citep{Price2018} evaluated only node-node interactions up to order $p=2$ (up to quadrupole moments) with leaf nodes as a target, reducing the number of long-range interactions by the mean number of particles in a leaf node (third row of \cref{fig:summarymmmethods}). This was found to give a performance gain of around ten compared to a standard tree code, even though its theoretical complexity is still $\mathcal{O}(N\log N)$. Other implementations, such as \cite{Fujiwara_1998}, cite FMM with complexity $\mathcal{O}(N \log N)$.

Such flavours are faster than standard tree codes and use \cref{eq:dmkdeffmm}. However, they do not guarantee that forces or torques are equal and opposite, as their paths are not identical during the expansion. This results in non-conservation of linear and angular momentum and hence centre-of-mass and angular momentum drift, respectively, in simulations. Exact conservation of linear momentum is possible in the standard FMM algorithm as it guarantees symmetric interaction paths between nodes \citep{Dehnen2000,Dehnen2002} (hereafter \citetalias{Dehnen2000} and \citetalias{Dehnen2002}). The algorithmic complexity $\mathcal{O}(N)$ derives from the implementation (fourth row of \cref{fig:summarymmmethods}).

\subsection{Angular momentum conservation} \label{sec:Angmomcons}
\citet{Marcello2017} (hereafter \citetalias{Marcello2017}) showed that one can also achieve exact angular momentum conservation. While the symmetric node-node interaction guarantees equal and opposite forces, one can add a small correction to ensure that the net torque also vanishes. In general, this requires a correction term involving the next order moments, in our case the octupole moments $Q^B_3$ (i.e. $Q^B_{xxx}  = \sum_j m_j b^x_j b^x_j b^x_j$ etc.).
This involves computing and storing the 10 extra independent quantities required for the octupole moments as well as the existing 6 quadrupole moments.

We apply the force correction from equation~19 of \citetalias{Marcello2017} which in our notation is
\begin{equation}
\bm{F}_{\rm corr} = \frac{G}{3!}\, \bm{D}'_4 \cdot \left(m_B\, \bm{Q}_3^A - m_A\, \bm{Q}_3^B\right),
\end{equation}
where $\bm{D}'_4$ is the fourth derivative of the Green's function with the terms parallel to $\bm{r}$ removed (see \citetalias{Marcello2017}) and $m_A \equiv Q_0^A \equiv \sum_j m_j$ is the mass of the node\footnote{Our equation also has a sign difference compared to Marcello because of our definition of $g(r)$ as $1/\vert r \vert$ not $-1/\vert r \vert$. Also, the prefactor of the corrective force in Eqs. (15) and (19) of \citetalias{Marcello2017} should be 
1/3! rather than $1/2$, consistent with Marcello's general expression (Eq. 38 in that paper) at $P=3$ ($p=2$ in our notation).}. The correction to the accelerations on each node is then the force divided by the mass of the respective node.  Since the corrective force is antisymmetric when swapping $A\to B$, linear momentum remains conserved by the scheme. As noted by \citetalias{Marcello2017}, the torque correction results in a slight loss of force accuracy, but this is small since the correction is already smaller than the truncation error of the expansion order.

The \citetalias{Marcello2017} correction requires moment expansion about the node centre of mass, so that the dipole terms vanish. Applying a constant force to an expansion centre other than the centre of mass creates non-zero torque perpendicular to the line connecting these two centres. This torque is not aligned with the interaction direction is not easily corrected. \citet{Korobkin2021} give a more general form, forcing the alignment to the direction of interaction with a projection operator, and using another Taylor expansion on the unit vector of the interaction direction on top of the multipole expansion. While their paper suggests a general approach to conserve angular momentum with any centre of expansion, in practice we found it to give poor accuracy in our tests ($p=2$ and correction terms up to octupoles). Hence, we expand about the centre of mass.

\section{Fast multipole method in {\sc Phantom}} \label{sec:num_impl}

\subsection{{\sc Phantom}'s tree and multipole acceptance criterion}

Despite the conservation advantage of the FMM over its siblings, implementing this method in a modern parallel hydrodynamical code is non-trivial. Multipole methods in astrophysics are always associated with hierarchical data structures like trees (kd-tree, binary tree, or oct-tree). In SPH, one gains by also using these structures to search for neighbours. A tree walk can check, level by level in a top-down fashion, whether a node in the tree could be a potential neighbour by satisfying a locality criterion. Multipole methods use a similar tree walk with slightly different acceptance criteria. 

In \textsc{Phantom}, the multipole acceptance criterion (MAC) is expressed as (\citetalias{Dehnen2000,Dehnen2014})
\begin{equation}
    \theta = \frac{l_\mathrm{tgt}+l_\mathrm{src}}{d} < \theta_\mathrm{crit},
\end{equation}
where $l_\mathrm{tgt}$ and $l_\mathrm{src}$ are respectively the size of the target and source node, $d$ is the distance between the two nodes and $\theta_\mathrm{crit}$ is the critical opening angle. The node size is defined as the largest distance between a particle of the node and its centre \citep[see: ][]{Price2018}. The MAC allows nodes to interact if they are well-separated, meaning that their opening angle is below the critical value. This criterion, along with the expansion order, controls the accuracy of the computation. The default value of $\theta_\mathrm{crit}$ is $0.5$ in {\sc Phantom}. This MAC is already well-suited to find the FMM symmetric interactions between target and source nodes. However, in addition to this MAC, we check a secondary acceptance criterion given by 
\begin{equation}
    d > [l_\mathrm{tgt}+l_\mathrm{src}+ R_\mathrm{kern}\mathrm{max}(h_{\mathrm{max_\mathrm{tgt}}},h_{\mathrm{max_\mathrm{src}}}) ],
\end{equation}
where $h_{\mathrm{max_\mathrm{tgt}}}$ and $h_{\mathrm{max_\mathrm{src}}}$ are respectively the maximum smoothing length inside the target and source node, and $R_\mathrm{kern}$ is the dimensionless cutoff radius of the smoothing kernel. This latter criterion ensures that every neighbour inside the smoothing kernel of the targeted particles will be handled individually to compute SPH forces and, more importantly here, direct softened gravitational forces \citep{Price2018}. This enables implementation of adaptive gravitational softening with FMM without modifying the Green's function presented in \cref{sec:multipolemethod}. We did not use improved MAC as described in \citetalias{Dehnen2014} as our preliminary results show no better accuracy per computational cost at our expansion order.

{\sc Phantom} uses a kd-tree to find neighbours and compute self-gravity. This choice was originally justified as follows: First, the construction of a kd-tree can be parallelised using the \citet{Gafton2011} algorithm with minimal memory footprint. Splitting nodes at their centre of mass minimises unbalanced situations when parts of the tree are deeper than others in the case of inhomogeneous and/or anisotropic distributions like accretion discs. The result is fast neighbour searches and gravity calculations. However, splitting nodes on the centre of mass can generate elongated node shapes in strong density gradients, resulting in larger minimal enclosing spheres for mass moments and node intersections. These larger spheres generate differences in the node size distributions on different levels of the tree. Mass moments are then computed over largely empty volumes. These two effects combined give lower force accuracy in the self-gravity in strong density gradients. Hence, we modified the tree build to use a geometric split along the longest axis. This avoids the creation of elongated nodes. However, we had to keep the centre of mass as the centre of expansion as explained in \cref{sec:Angmomcons}. This also allows us to skip the calculation of the dipole mass moments that vanish at this position. 

We only store the quadrupoles and the octupoles (for angular momentum corrections) for each internal node and leaf node. According to {\sc Phantom} default value, the leaf nodes can hold a maximum of $10$ particles. These mass moments are computed using \cref{eq:massmom} for each node during the tree-building step from the root to the leaf nodes in conjunction with the particle-centre distance scan needed to build the minimal enclosing sphere around the node centre of expansion used to walk the tree. That way, we use the already computed distances to construct the mass moments. This method showed better performance than a back-propagation from leaf nodes to the root using \cref{eq:massmomtrans}, as the node sizes cannot be extrapolated from its children as would be the case in a geometrically centred node. An alternative would have been to use an approximation of the minimal enclosing sphere radius. However, this would produce inefficient neighbour searches and potentially larger errors in the self-gravity calculation.

\subsection{Parallel-split dual tree traversal}

The key fix needed to symmetrise the previous scheme is the dual tree traversal presented in \citetalias{Dehnen2000}. A difficulty is that embarrassingly parallel implementation of this is not possible because the algorithm is recursive. \cite{Taura_2012} and \citetalias{Dehnen2014} used task-based parallelisation to share the load of the top-down search between multiple threads on the CPU. While this method gave satisfactory performance on multi-core architectures, we found it to be incompatible with {\sc Phantom}'s overall parallelisation strategy. Moreover, modern accelerators like GPUs are not well-suited for task-based parallelism. The few attempts to port parts of the code on GPU in the community, and the close development proximity of the Shamrock GPU-native code \citep{David--cleris_2025} led us to desire a future-proof implementation.

Following the definition given in \cite{Curtin_2013}, a dual tree traversal algorithm recursively pairs nodes of two trees, the query (target) and the reference (source) tree, to test a certain ``interaction'' function (in our case the well-separatedness of nodes). In Dehnen's traversal, the query and reference tree are identical and equal to the whole tree structure. Instead, we chose to split the query tree into every branch (a specific subtree of the whole one) that connects a leaf node and the tree root. Effectively, this splitting allows us to assign an independent dual tree traversal to each thread of a multi-core architecture, making the whole algorithm embarrassingly parallel.

This strategy implies a duplication of the interaction tests made on shared parts of different branches of the tree. But since each thread is independent, the duplication does not double-count interactions. The resulting interaction list produced by this algorithm is equivalent to what would give a unsplit dual tree traversal. The splitting strategy also implies that we lose the expected $\mathcal{O}(N)$ scaling. The complexity of one branch scan is given by the tree depth, $\mathcal{O}(\log N)$. As the algorithm performs $N_{\rm leaf}\propto N$ scans, the final complexity falls back into a $\mathcal{O}(N\log N)$ complexity. However, we gain back momentum conservation with a trivially parallel tree traversal.

\begin{algorithm}
\caption{Parallel-split dual tree traversal}\label{alg:DTT}
\KwData{
        $\{n_i\}_{i \in [0,n)}$ tree leaf nodes list
        }
\KwResult{$\{\mathrm{d}M_{n_i}\}_{i \in [0,n)}$ gravitational field tensors}
\BlankLine
\tcp{Main loop on the leaf nodes}
\For{$n_i$ {\rm in parallel}}{
    ancestors = get\_leaf\_branch($n_i$)\\
    \tcp{ancestor(1) is always the root}
    push onto stack (ancestors(1), root)\\
    \While {\rm stack not empty}{
        ($a_i,\, \mathrm{b}$) = pop stack\\
        \If{$a_i == b$}{
            stackit = true\\
        }
        \Else{
            \If{{\rm MAC}($a_i$,\,$b$)}{
                node\_node\_interaction($\mathrm{d}M_{a_i}$,$a_i$,$b$)\\
            }
            \Else{
                stackit = true\\
            }
        }
        \If{$\mathrm{stackit}$}{
            \If{$b \; \mathrm{is \; leaf}$}{
                \If{$a_i\; \mathrm{is \; leaf}$}{
                    $\mathrm{cache\_neighbours}(a_i,b)$\\
                }    
                \Else{
                    push onto stack ($a_{i+1}$,$b$)\\
                }
            }
            \Else{ 
                \If{$a_i\; \mathrm{is \; leaf}$}{
                    push onto stack ($a_i$,$b_l$)\\
                    push onto stack ($a_i$,$b_r$)\\
                }    
                \Else{
                    push onto stack ($a_{i+1}$,$b_l$)\\
                    push onto stack ($a_{i+1}$,$b_r$)\\
                } 
            }
    }
    }  
    \For {$a_i$ {\rm in ancestors}}{
        propagate\_to\_next\_ancestor($\mathrm{d}M_{n_i},\mathrm{d}M_{a_i},a_i,a_{i+1}$)\\
    }  
}
\end{algorithm}

\Cref{alg:DTT} describes our new implementation of a dual tree traversal in pseudo-code that we refer to hereafter as the Parallel-split dual tree traversal. The parallel loop in this algorithm is performed on the leaf nodes of the tree as follows:
\begin{enumerate}
    \item The list of ancestor nodes of a leaf is retrieved up to the root. This constitutes the query tree. To start the dual tree walk, the farthest ancestor of the leaf (which is always the root) and the root are pushed onto a stack of interactions. 
    \item The main walk is done by popping the last interaction off the stack and testing its opening angle against the acceptance criterion. If the MAC passes, the two nodes are well separated, the node-node interaction is processed, and the gravitational field tensors are stored in the ancestor buffer. If not well-separated, multiple outcomes are possible depending on the nature of the nodes:
    \begin{enumerate}
        \item If the two nodes are both leaf nodes, the tree cannot be walked any more. Therefore, all interactions between these two nodes should be computed particle by particle. Hence, the source node particles are stored in a cache that will be used later to compute these short-range interactions.
        \item If the two nodes are not leaf nodes, the next ancestor node (from the query tree) and the two children of the source node are pushed onto the stack. Pushing only the next ancestor on the target side guarantees to compute only the interactions needed for the targeted leaf without losing the symmetry. Opening both nodes at the same time also helps to minimise the error by limiting the size difference between nodes \citep{Springel2021}.
        \item The intermediate case can be problematic, where one leaf node is shallow in the tree. Following \citetalias{Dehnen2000} and \citet{Springel2021}, these interactions must be handled particle by particle. However, due to the symmetry between target and source nodes, this can involve a large number of direct sum calculations, ruining the performance benefit of the FMM. We avoided this by allowing shallow leaf nodes to be tested with deeper nodes in the tree. This lowering method can be applied to both the target and source sides of the interaction without breaking the symmetry.
    \end{enumerate}
    \item This process is repeated until the stack is empty.
    \item Lastly, we loop over the list of ancestor nodes to the leaf, from top to bottom. During this loop, the gravitational field tensors are summed and translated from the parent to children until reaching the leaf node using \cref{eq:gravmtranslat}. After this step, the gravitational field tensors are translated onto individual particle positions to get the long-range force. The cached neighbour list is then be for short-range interactions and SPH forces.
\end{enumerate}

\begin{algorithm}
\caption{Query branch pruning}\label{alg:pruning}
\KwData{\{ancestors\} Query branch }
\KwResult{\{stack\} Pruned traversal stack}
\BlankLine
\tcp{Loop over the ancestors (leaf to root) and find the first fully cached}
\For{$a_i$ {\rm in ancestors}}{
    \If{$a_i$ {\rm interaction list is fully cached}}{
        \For{$b$ {\rm in interaction list}}{
            push onto stack($a_{i+1}$, $b_l$)\\
            push onto stack($a_{i+1}$, $b_r$)\\
        }
        break the loop\\
    }
}
\end{algorithm}

\begin{algorithm}
\caption{Cache writing}\label{alg:cachewrite}
\KwData{$\{a_i\}$ ancestor node $i$}
\KwResult{$\{\mathrm{d}M_{a_i}^c,l_{a_i}^c\}$ Cached node properties}
\BlankLine
\tcp{Cache writing barrier}
\tcp{${\rm cachestate}_{a_i}$ is the public mutex}
\tcp{${\rm cachestate}$ is the private capture}
\tcp{{\bf atomic capture} capture the state of ${\rm cachestate}_{a_i}$ and set it to ``claimed'' atomically if untouched}
{\bf atomic capture}(${\rm cachestate}_{a_i}$,${\rm cachestate}$,``claimed'')\\
\If{${\rm cachestate}$ == {\rm ``untouched''}}{
    $\mathrm{d}M_{a_i}^c = \mathrm{d}M_{a_i}$\\
    {\bf atomic write}(${\rm cachestate}_{a_i}$,``${\rm d}M$-cached'')\\
    \If{${\rm length}(l_{a_i}) < {\rm threshold}$}{
        $l_{a_i}^c = l_{a_i}$\\
        {\bf atomic write}(${\rm cachestate}_{a_i}$,``fully-cached'')\\
    }
 }
\Else{
    \If{{\rm cachestate is }${\rm d}M{\rm-cached}$}{
        $\mathrm{d}M_{a_i} = \mathrm{d}M_{a_i}^c$\\
    }
}
\end{algorithm}

\subsection{Caching strategy}

This algorithm makes the parallel scheduling trivial and therefore maximises the parallel efficiency of the code. However, the extra interactions from the upper query branches (tree) linked to the leaf nodes can lead to slightly lower performance than the previous non-symmetric scheme. This extra cost can be avoided by using a caching strategy to efficiently prune the traversal of most of the tree traversals. All query branches (trees) share a common prefix of internal nodes that start at the root. Our caching strategy consists of storing the gravitational tensor fields and the neighbour lists (non-well-separated interactions) for each of these shared internal nodes. That way, when a thread starts to work on a new branch, it will first check for already cached nodes within it and start the tree traversal at the deepest cached node of the branch. On paper, this optimisation is able to recover the asymptotic $\mathcal{O}(N)$ complexity of the original FMM algorithm. Indeed, if threads work cooperatively, the uncached depth of a query branch is $\mathcal{O}(1)$ compared to the full $\mathcal{O}(\log{N})$ depth.

Implementation of this caching strategy needs to be done carefully to avoid race conditions and inefficient memory access. We opted for a shared cache array to store both the gravitational tensor fields ${\rm d}M_a$ and the list of neighbours $l_a$ of all the internal nodes of our tree. The main part of this caching process is a mutex lock and atomic operations that prevent race conditions between threads wanting to write to or read from the same cache.

Each internal node is associated with a status flag (``cachestate''). This first acts as a barrier for writing operations. Then it is also used to authorise a cache read operation. During the tree build, the flag is initialised to an ``uncached'' state for each node of the tree. When a thread finishes its full traversal, it will loop over its internal node and try to claim its ``untouched'' state with an atomic update. If the returned flag is still ``untouched'', it is the first one to reach its cache. The atomic update modifies the public value of this mutex to ``claimed'', meaning that no other thread will try to cache it. This is done synchronously thanks to the atomic, avoiding races between threads on the write operation. After this barrier, writing the values of ${\rm d}M_a$ and $l_a$ can be done safely on the shared cache. To flag that the cache has been written, the writing thread ends the operation by setting the ``cached'' state to the node's mutex to allow other threads to use it. This is also done atomically to avoid mutex corruption.

The above is essentially a lock, but to avoid congestion and waiting from one thread to another, we let threads race to win the cache write. That way, we do not lose parallel efficiency while saving time when a thread restarts on an already cached branch.  

The caching strategy can be broken down into two separate operations, the pruning phase and the cache writing phase. These phases are respectively written in pseudocode in \cref{alg:pruning} and \cref{alg:cachewrite}. Pruning is done just before entering the traversal loop, and cache writing inside the propagation loop. A few other implementation details should be mentioned. The number of non-well-separated nodes is theoretically $\mathcal{O}(1)$; however, storing the interaction list for all the nodes can rapidly add up in the cache, making it impractically large. Hence, we put a threshold on the size of this list. We found an optimal value of $1024$ for the size cap. Over this threshold, interaction list caching is skipped, although ${\rm d}M_a$ is always stored. Therefore, we split the ``cached'' mutex state into two to assess the two kinds of caching(``${\rm d}M$-cached'' and ``fully-cached''). That way, we can still skip the gravitational tensor field computation while doing the node interaction test. The following section shows the performance improvements this strategy offers.

\begin{figure*}
    \centering
    \includegraphics[width=0.95\linewidth]{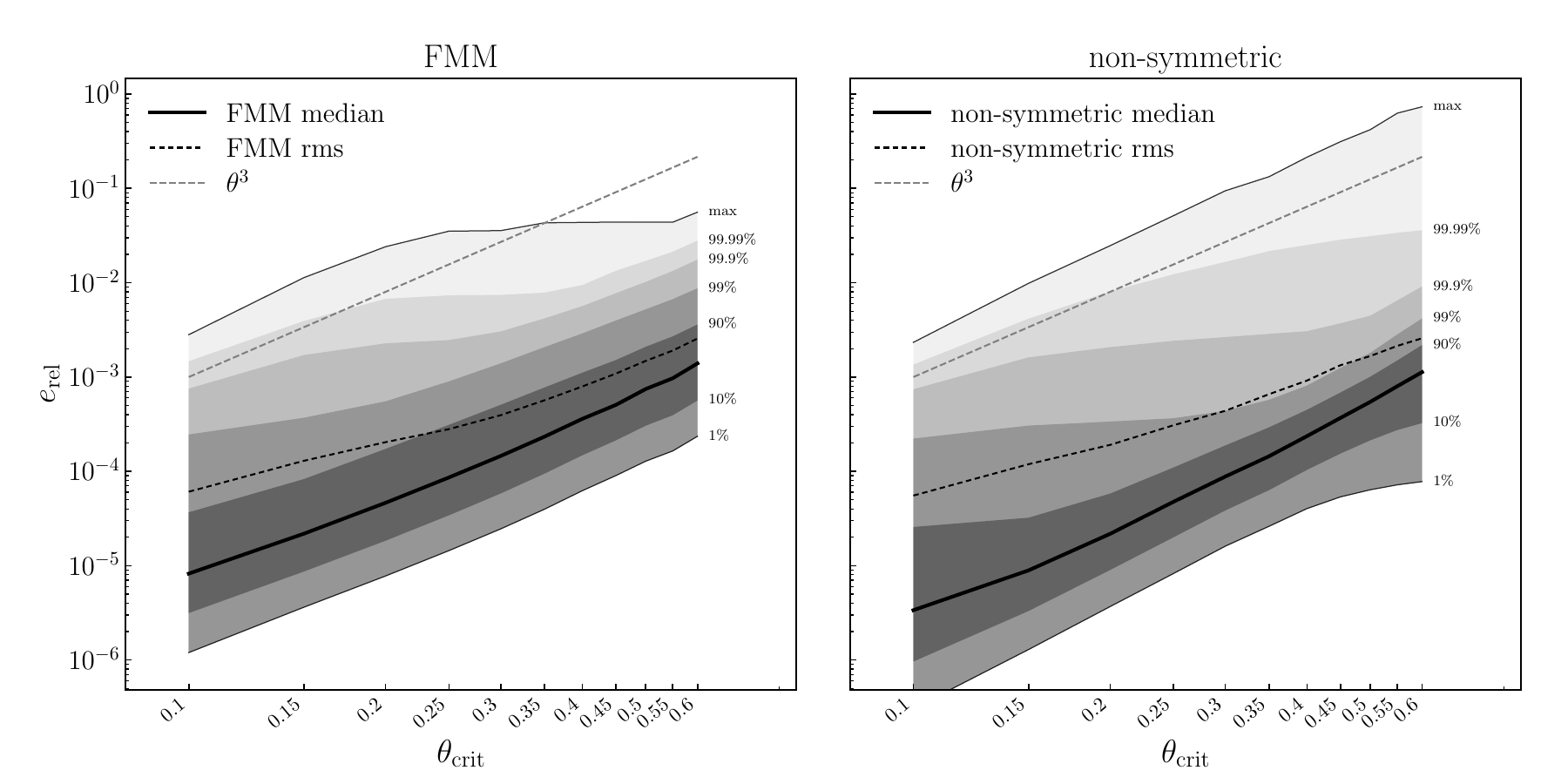}
    \includegraphics[width=0.95\linewidth]{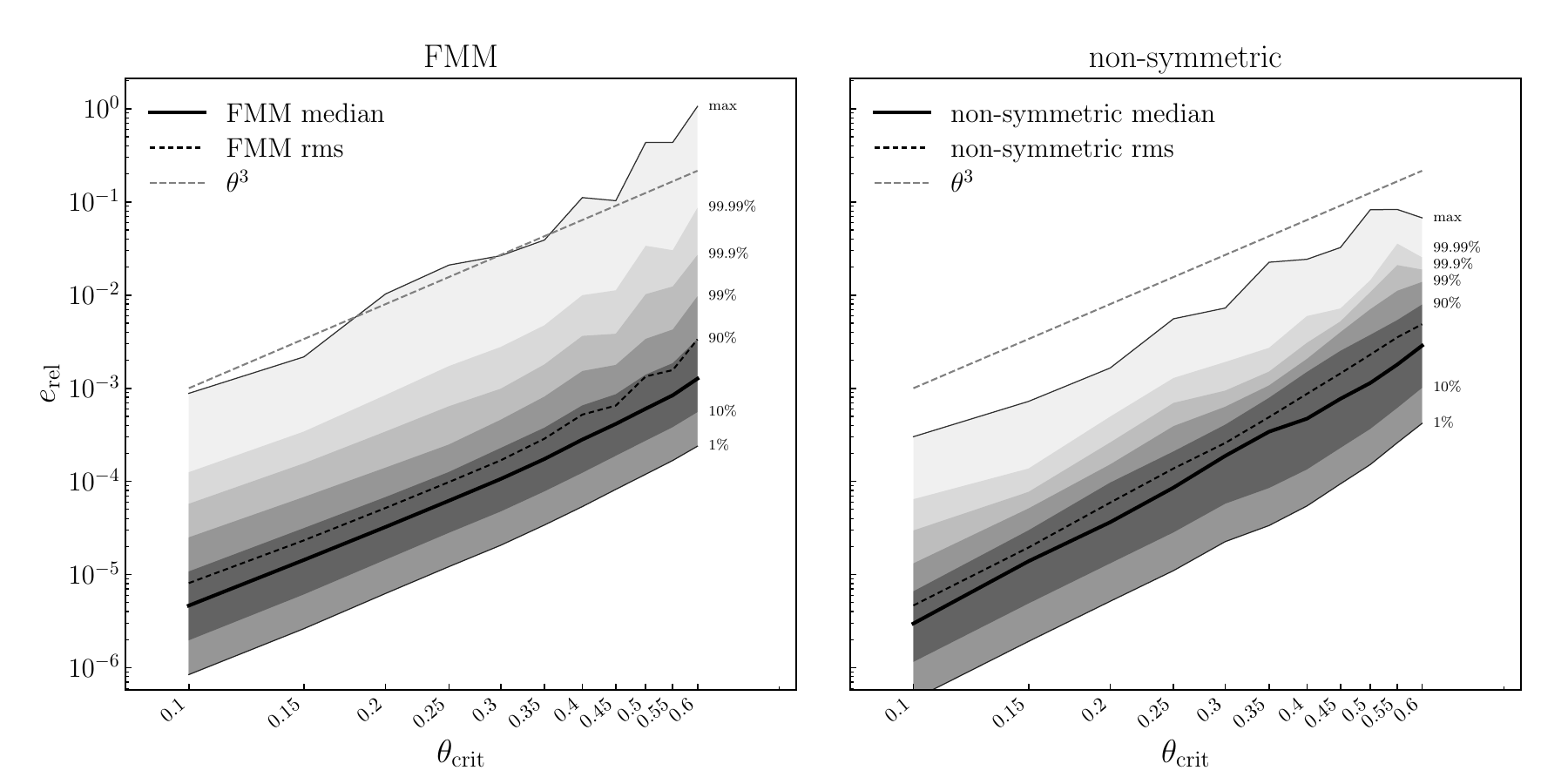}
    \caption{Median relative force error (compared to a direct sum) versus maximum opening angle computed in a Plummer sphere  (top) and a homogeneous sphere (bottom) of $3\times10^6$ equal-mass particles, comparing the new FMM (left) to the original {\sc Phantom} self-gravity (right). Shading around each curve represents percentiles between the maximum error (upper) and the first percentile (lower).
    }
    \label{fig:sphere_err}
\end{figure*}

\section{Results}\label{sec:results}
We present tests and benchmarks of our FMM implementation to measure its precision and numerical performance. Tests were performed using the Grenoble University local cluster using AMD EPYC 9654 nodes using 48 cores, except for the common-envelope evolution benchmark (Sect. \ref{subsec:CEE}), which used 48 cores on a single AMD EPYC 9754 processor in the internal computing cluster of the Heidelberg Institute for Theoretical Studies.

\begin{figure}
    \centering
    \includegraphics[width=\linewidth]{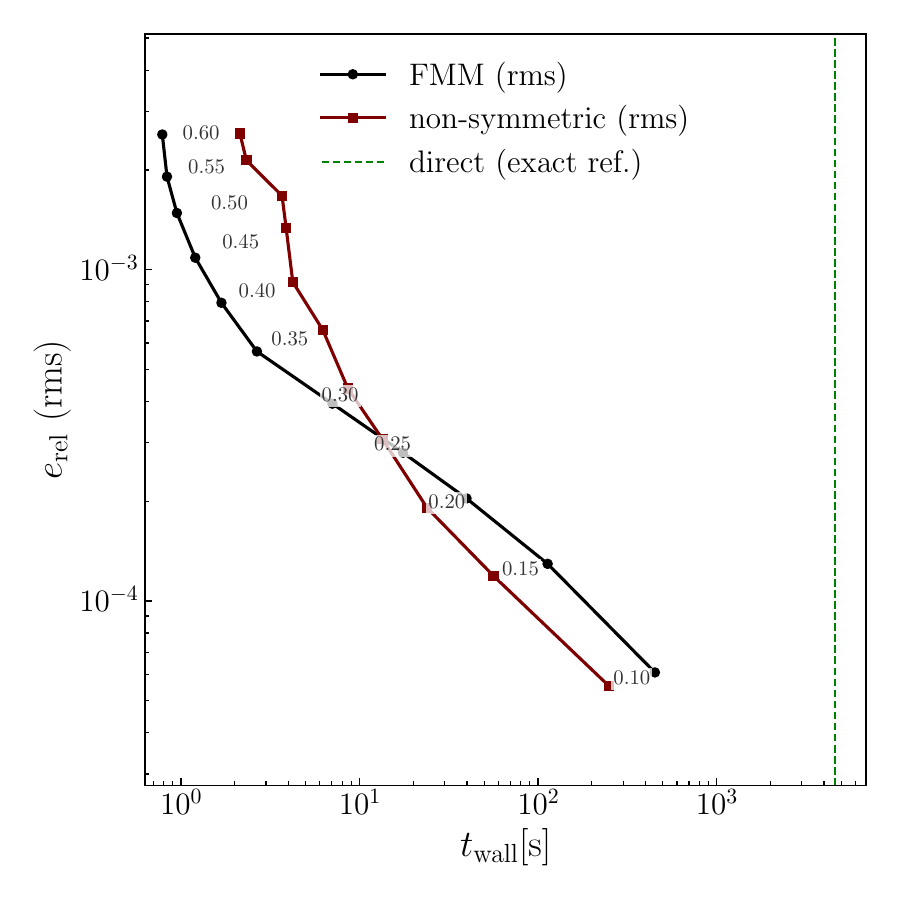}
    \caption{Accuracy versus cost. Relative RMS error versus the wall time (in seconds) spent computing gravitational forces in a Plummer sphere of $3\times10^6$ particles for $\theta_{\rm crit}$ ranging from $0.1$ and $0.6$. The original {\sc Phantom} self-gravity is colored in red while the new FMM is in black. The reference time of the direct algorithm is plotted in green and is unrelated to the critical opening angle.}
    \label{fig:plummer_perf1}
\end{figure}

\subsection{Plummer and homogeneous sphere}
First, we computed gravitational accelerations for particles placed in a Plummer sphere \citep{Plummer1911}, comparing the previous and new algorithms. The previous scheme uses non-symmetric node-node interactions with only target leaf node in combination with the previous centre-of-mass-split tree, while our new implementation use symmetric node-node interactions and the new geometric splitting. We generated a randomly drawn distribution of $3\times10^6$ equal-mass particles placed (as described in \citealt{Price2007} following \citealt{Aarseth_1977}) to initialise and maintain the density profile of a Plummer sphere
\begin{equation}
    \rho(r) = \frac{3Ma^2}{4\pi(a^2+r^2)^{5/2}},
\end{equation}
where $M=1$ (code units) is the total mass and $a=1$ (code units) is the characteristic radius, also called Plummer radius. We computed gravitational forces with both methods for eleven maximum opening-angle criteria, $\theta_{\mathrm{crit}}\in [0.1,0.6]$, with $0.05$ step increments. We compared to forces computed with a direct sum. 

\Cref{fig:sphere_err} (top) shows the median relative error ($e_{\rm rel} \equiv 1/N\sum_i|a_i - a_{i,{\rm direct sum}}|/|a_{i,{\rm direct sum}}|$) over the whole $\theta_{\mathrm{crit}}$ range, comparing our new FMM (left) to the previous scheme (right). The shadings represent the spread of errors between the maximum and the tenth percentile. We also show percentiles near the maximum and minimum to illustrate the error distribution. The two methods give similar results for small $\theta_{\mathrm{crit}}$. FMM gives $2.4$ to $1.2$ times less accurate median values compared to the previous scheme from $\theta_{\mathrm{crit}}=0.1$ to the end of the range. However, the new FMM produces lower maximum errors than the previous method at larger $\theta_{\mathrm{crit}}$. Specifically, the FMM RMS error value is about $20\%$ lower at large $\theta_{\mathrm{crit}}$ than the previous method.

The lower panel of \cref{fig:sphere_err} shows the corresponding results for a homogeneous density sphere. In this case, the FMM gives a better median error ($20$ to $50\%$ lower) with $\theta_\mathrm{crit}$ between $0.2$ and $0.6$. The FMM gives higher maximum and high percentiles ($>99\%$) errors than the previous scheme. However, these significantly higher errors only impact $\sim3\,000$ particles in the whole sphere. We will discuss further on this trend in a later section that is a combination of our new angular momentum conservation patch and the relative error computation of this test.

\Cref{fig:plummer_perf1} shows the accuracy versus cost comparison, plotting the RMS relative errors measured on our Plummer sphere test for FMM (black line) and the previous scheme (red line) as a function of the wall time for the different opening angles. The direct sum algorithm is also shown as a reference.$\theta_\mathrm{crit} = 0.25$ is a turning point where the computational efficiency of the new FMM starts to pay off. At a critical angle of $0.5$, which is the default option in the code, our FMM is four times faster than the previous scheme for roughly the same RMS value. Hence, we achieve better performance improvement at fixed accuracy.

\begin{figure}
    \centering
    \includegraphics[width=\linewidth]{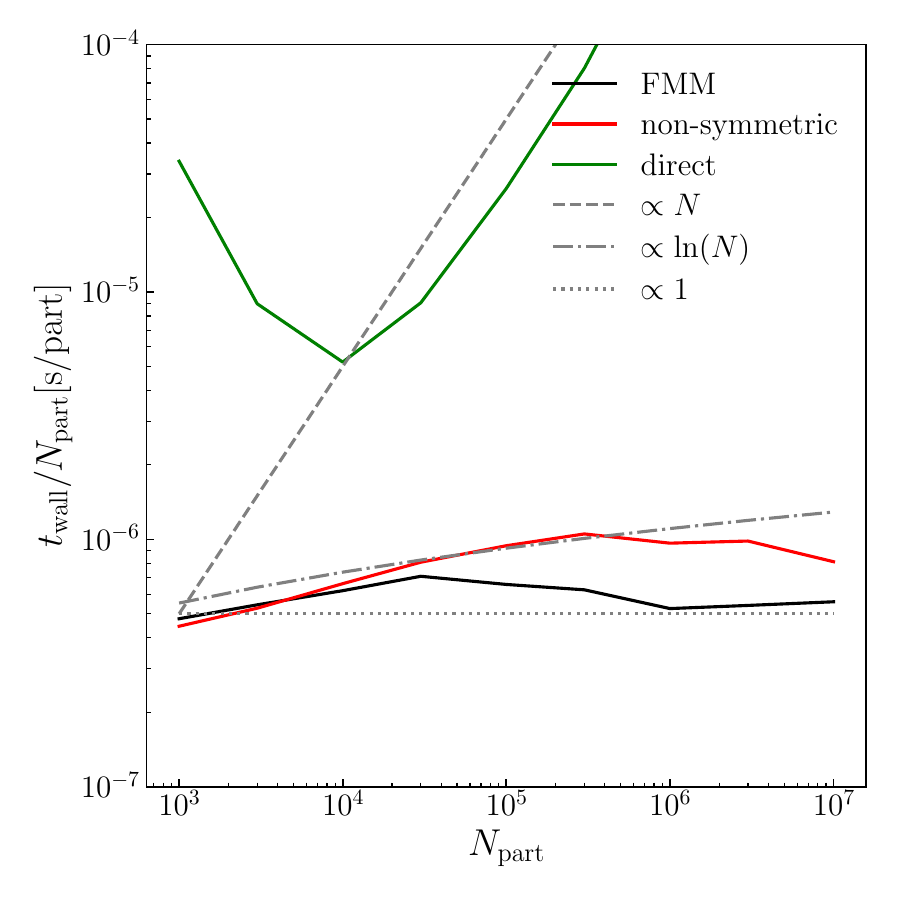}
  \caption{Wall time divided by the number of particles (in seconds) spent computing gravitational forces in a Plummer sphere with $10^3$ to $10^7$ particles with $\theta_\mathrm{crit}=0.5$. The two multipole methods, {the previous scheme} in red and FMM in black, are compared to the direct sum algorithm in green.}
    \label{fig:plummer_perf2}
\end{figure}

\Cref{fig:plummer_perf2} shows the wall time for computing all gravitational interactions in the Plummer sphere with $\theta_\mathrm{crit}=0.5\,$, comparing direct sum (green line), the previous scheme (red line) and our new FMM (black line), divided by the number of particles, ranging from $10^3$ to $10^7$. The direct method can be seen to follow an $\mathcal{O}(N^2)$ trend, as expected. Our new implementation (\cref{alg:DTT,alg:cachewrite,alg:pruning}) shows a roughly constant value across the whole range of the number of particles. This shows that our caching strategy is able to recover the theoretical $\mathcal{O}(N)$ complexity of the FMM algorithm. By contrast, our previous scheme can be seen to follow the $\log(N)$ trend, even though the wall time per particle seems to flatten or even decrease for $N \gtrsim 10^6$ (see Discussion).

 \begin{figure*}
    \centering
    \includegraphics[width=0.9\textwidth]{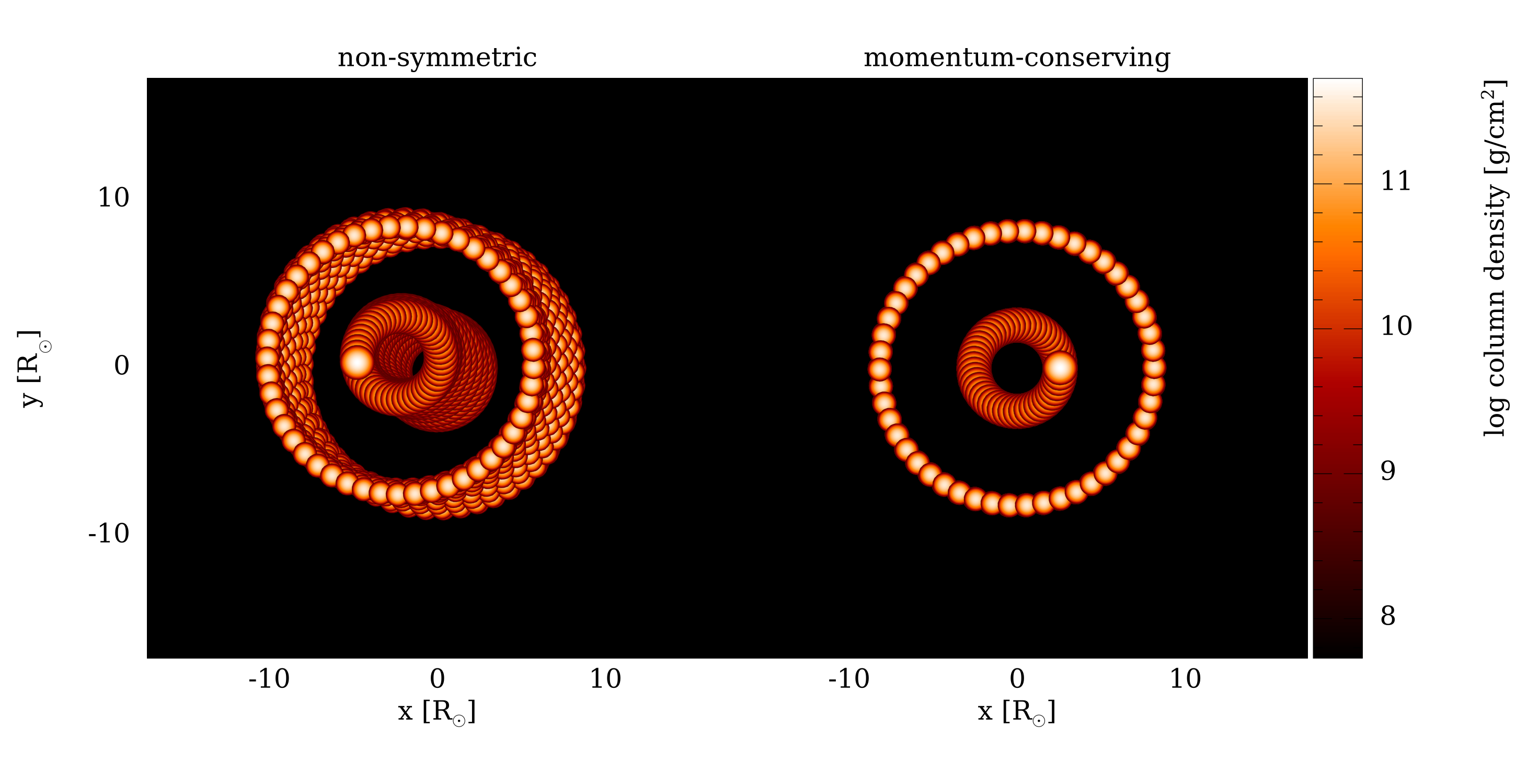}
    \vspace{-1cm}
    \caption{Twenty orbits of a binary system of polytropic stars. The render of the logarithmic column density of each timestep is stacked to construct the path of the binary components. With the previous scheme (left), a small but accumulative non-conservation error causes the orbit to drift with time, while this drift is removed with the new FMM implementation (right).}
    \label{fig:bin_orbits}
\end{figure*}

\subsection{Binary polytrope test}
We computed the evolution of a binary star modelled by two polytropes consisting of a primary star of $1 \,\Msun$ and $1 \,\Rsun$ and a companion of $0.314 \,\Msun$ and $0.680\,\Rsun$ (equal density), set up using the automated Relax-O-Matic$^{\rm TM}$ procedure in {\sc Phantom} \citep[Appendix C of][]{Lau2022a}. The two gaseous spheres are set orbiting around each other in a circular orbit with a semi-major axis equal to $0.05\,\au$. The mass resolution was set to $10^{-4}\,\Msun$ giving a total particle number around $1.3\times10^4$. We used an ideal gas equation of state with $\gamma=5/3$ to model the polytropes. During this test, the binary motion was integrated for 20 orbits with the previous scheme and our new FMM. 

\Cref{fig:bin_orbits} shows logarithmic column density renderings of the binary, stacked and displayed for both methods at intervals of 1\% of the orbital period. The plot shows the trajectory of both stars over the 20 orbits. The non-symmetric force evaluations computed with the previous scheme generate a visible drift of the binary barycentre (left panel) that is not present with the FMM version (right panel). With the previous scheme, the centre-of-mass can be seen to drift (to the upper left in the figure) by a significant fraction of the semi-major axis. By contrast, the binary trajectory stays stable during the whole simulation with our new FMM. The FMM calculation also shows a perfect orbital phase recovery, with the final position of the stars perfectly superimposed on their initial location. By comparison, the non-symmetric version does not conserve this property.

\begin{figure}
    \centering
    \includegraphics[width=0.95\linewidth]{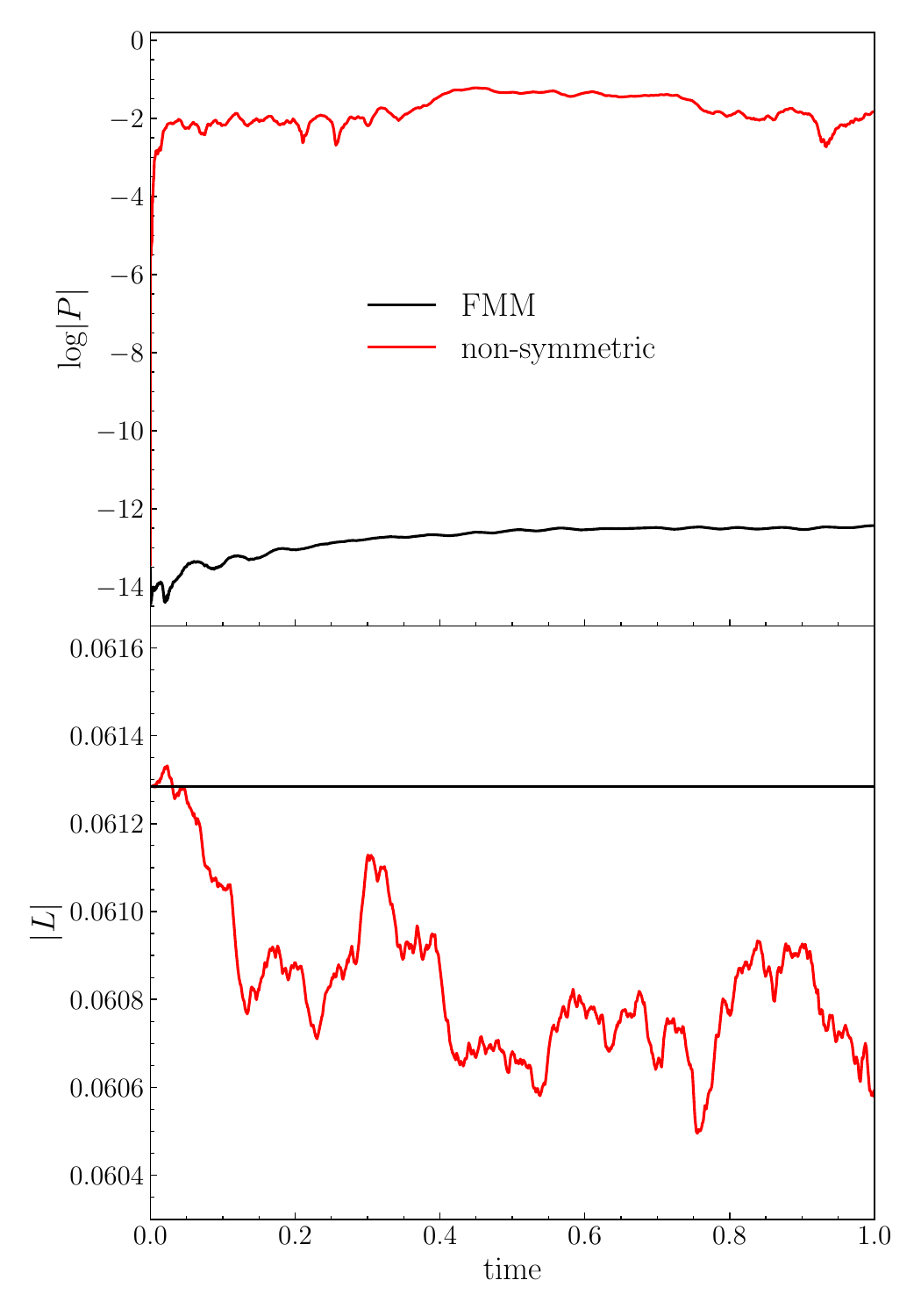}
  \caption{Evolution of the linear momentum (top) and angular momentum (bottom) of a binary system of polytropic stars over 20 orbits. The previous non-symmetric scheme and FMM are plotted respectively in red and black. The linear momentum in the FMM case is conserved up to machine precision over the whole simulation. Angular momentum conservation is also conserved up to machine precision thanks to the force correction.}
    \label{fig:bin_conservation}
\end{figure}

\Cref{fig:bin_conservation} shows the corresponding evolution in time of both linear (top) and angular (bottom) momentum of the two simulations reported in code units ($G=\Msun=\au=1$). The linear momentum deviation (relative to zero; top panel) never exceeds a value of a few $10^{-13}$ with FMM, consistent with conservation to machine precision, while the previous method rises to $10^{-2}$ over the 20 orbital periods simulated. With \citetalias{Marcello2017}'s force correction added to the FMM, we obtain conservation of the angular momentum to machine precision (black line in bottom panel). By contrast, the previous version shows angular momentum drift over the simulation with relative fluctuations of order $10^{-4}$ (red line in lower panel).

\subsection{Common-envelope evolution}
\label{subsec:CEE}
To demonstrate the capabilities of the FMM algorithm on a more complex, self-gravitating astrophysical problem, we compare the previous {\sc Phantom} scheme with the new FMM in a simulation of common-envelope evolution. Common-envelope evolution is a phase in the evolution of binary stars where one star (the donor) engulfs its much more compact companion (the accretor) as it expands as a giant star \citep{Paczynski1976,Ivanova2013,Roepke+DeMarco2023,Schneider2025}. The companion and the giant star's core orbit each other inside the tenuous giant envelope, where gravitational drag causes them to spiral inwards over several orbits. Some or all of the giant envelope is ejected in this process. Common-envelope evolution has been simulated extensively with \textsc{Phantom} \citep{Iaconi2017,Iaconi2019,Reichardt2019,Reichardt2020,Lau2022a,Lau2022b,Gonzalez-Bolivar2022,Gonzalez-Bolivar2024,Bermudez-Bustamante2024,Lau2025,Nibbs2025}. We use the same setup and method as \cite{Lau2022a} for the ``gas + radiation'' equation of state that assumes radiation to be in thermodynamic equilibrium with the gas and for the ``low-resolution'' case that only uses 50,000 SPH particles.

\begin{figure*}
    \centering
    \includegraphics[width=0.95\textwidth]{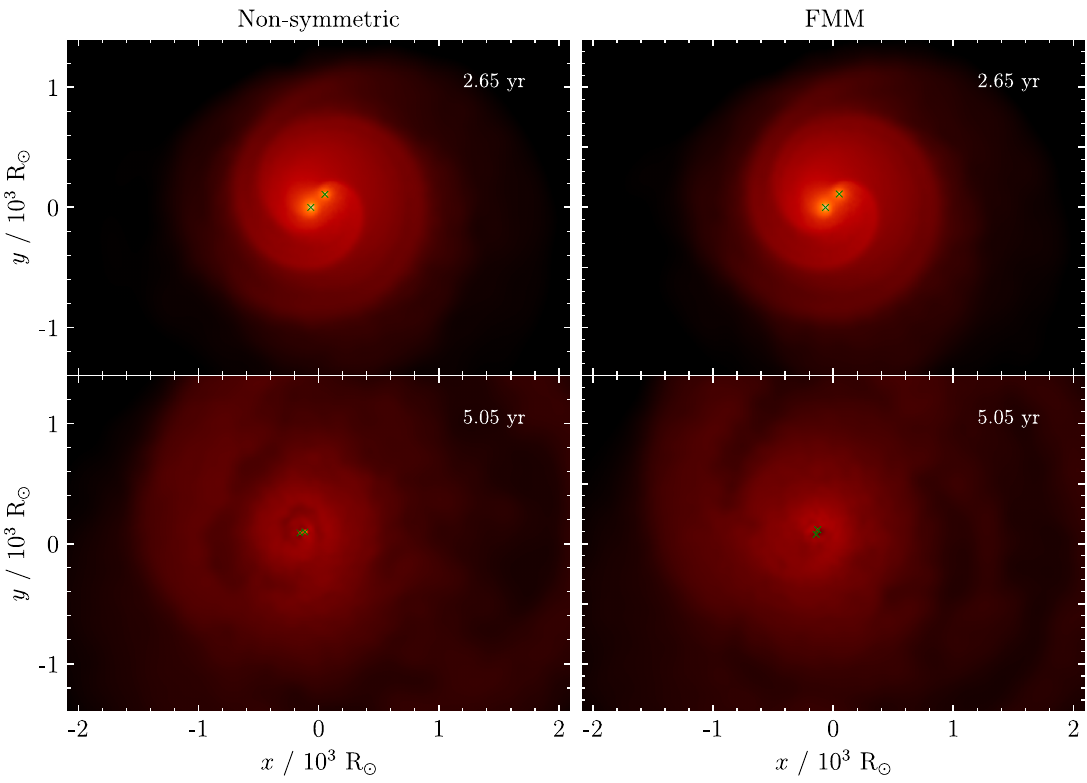}
    \caption{Density slices at $t=2.65,\ 5.05\ \mathrm{yr}$ in the orbital plane of a simulation of common-envelope evolution, comparing results obtained with the momentum-conserving method (right) with those obtained with the previous non-symmetric method (left). The green crosses show the positions of the point masses representing the stellar core and the companion star. The results are indistinguishable up to the end of the plunge-in where the ejecta becomes turbulent. Plot made using \textsc{Sarracen} \citep{Harris+23}.
    }
    \label{fig:render_CE}
\end{figure*}

\Cref{fig:render_CE} compares simulations performed with the previous scheme and with the FMM. The first row shows snapshots taken during the dynamical spiral-in phase of the stellar cores, which are visually indistinguishable between the two methods. The second row shows a phase in the late-time evolution after much of the original giant envelope has dispersed and some parts have been ejected. An exact match between the two methods is no longer possible as turbulent, convective flows have developed in the ejecta, exponentially amplifying small initial deviations.

\begin{figure}
    \centering
    \includegraphics[width=0.9\linewidth]{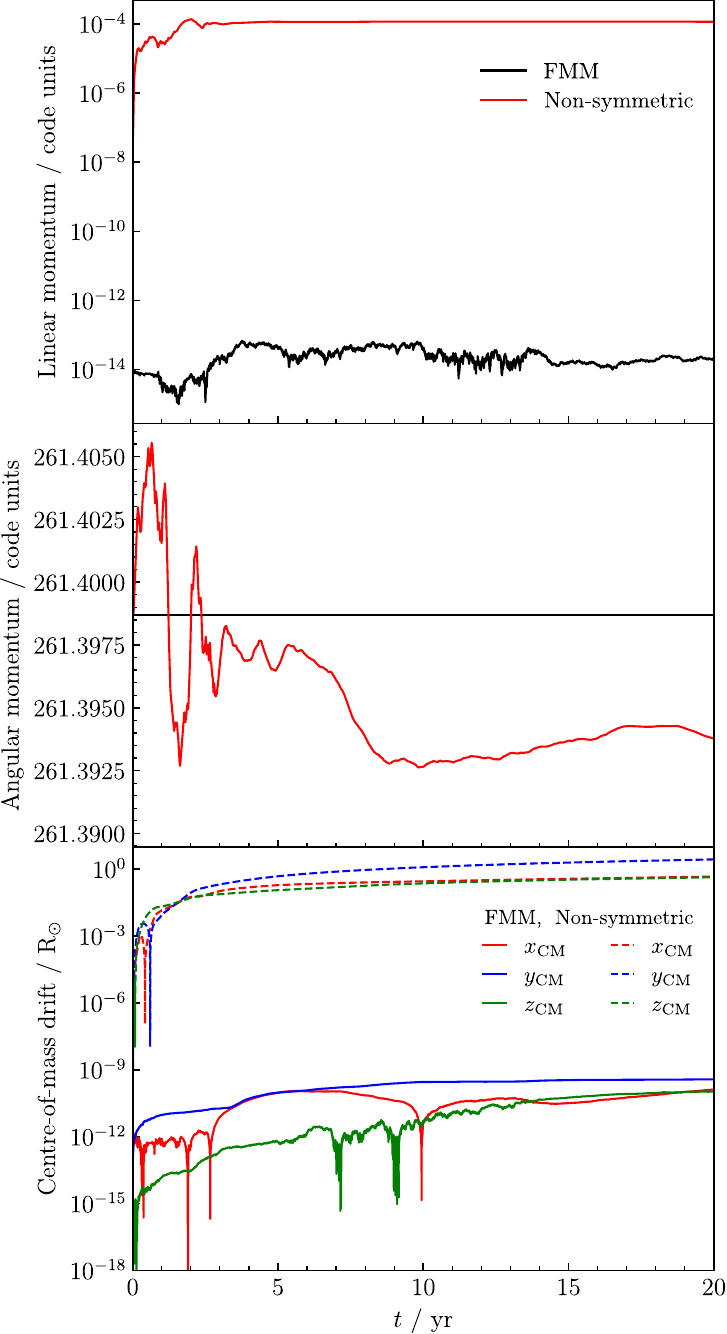}
    \caption{Similar to \cref{fig:bin_conservation}, but for the common-envelope simulation. Bottom panel compares the $x$, $y$, and $z$-centre-of-mass drifts between the FMM and the original method.}
    \label{fig:conservation_CE}
\end{figure}

\Cref{fig:conservation_CE} compares the evolution in total linear and angular momentum between the previous scheme and FMM. A similar improvement in linear momentum conservation is observed compared to the binary polytrope test. The simulation was set up to have zero linear momentum initially. With the previous scheme, total momentum is conserved to within $10^{-4}$ (in code units) and to within $10^{-13}$ with the FMM, the latter consistent with conservation to round-off error. There is no discernable difference in computational cost. With the new method, the centre-of-mass drift is restricted to within $\sim 10^{-9}\Rsun$, whereas it used to reach $\sim 1\Rsun$. \Cref{fig:conservation_CE} also shows that the new scheme conserves angular momentum to round-off error.

Conserving total angular and linear momentum is especially important for capturing the orbital dynamics after the initial fast plunge-in. In this phase, the secular orbital evolution is influenced by gravitational interaction with the remaining bound envelope, and the morphology and velocity of bipolar outflows and jets could be highly sensitive to small drifts in the barycentre.

\begin{figure}
    \centering
    \includegraphics[width=0.9\linewidth]{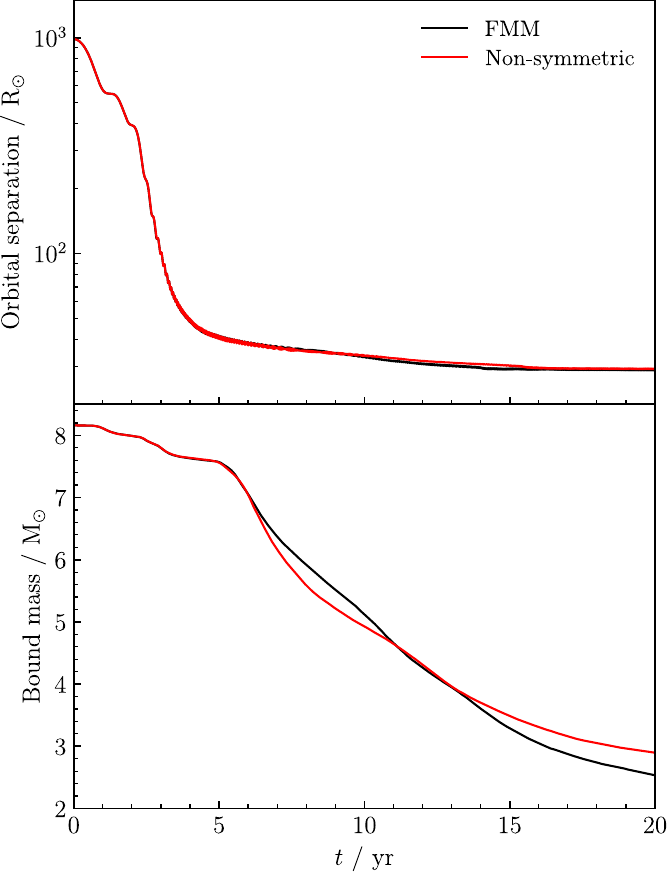}
    \caption{Comparison of key results between the FMM and original method for the common-envelope evolution simulation. Top panel: Separation between the companion and the donor star's core. Bottom panel: Total envelope mass that is bound, that is, possessing a negative sum of kinetic and potential energies.}
    \label{fig:ce_sep_boundmass}
\end{figure}

\Cref{fig:ce_sep_boundmass} compares the key results in a simulation of common-envelope evolution: the evolution in the amount of unbound envelope mass and the separation between the giant core and the companion. Due to excellent angular momentum conservation in both the {original method} and FMM, there are only minor differences in the evolution of orbital separation and almost no discernable difference up to the immediate end of the plunge-in phase ($t\approx 4~\yr$). There are larger differences ($\sim10\%$) observed in the amount of bound envelope mass after the plunge-in. As noted earlier, an exact match between the two simulations is not expected during this phase.

\section{Discussion}\label{sec:discuss}

Our tests show the advantages of symmetrising our previous self-gravity scheme. Conserving linear and angular momentum up to machine precision allows a better physical description of the studied system, as shown with the binary polytrope test (see~\cref{fig:bin_orbits}). The centre-of-mass drift discussed in \citet{Potashov2025} is absent. Better momentum conservation also improves the accuracy of other quantities, such as the orbital phase, recovered perfectly with this symmetrisation of the self-gravity. Linear momentum conservation is enough to recover this orbital phase. Adding the angular momentum conservation by applying the \citetalias{Marcello2017} force corrections may improve the accuracy in simulations where torque transfer is important, such as self-gravitating discs or tidal disruption events. We measured the cost of these force corrections as $\approx 5$--$10\%$ compared to the computational cost of a full global time step. We consider this extra cost sufficiently low to be worth using as a default option in the code. The self-gravity improvements build on the conservation properties of the SPH scheme, making the entire code linear and angular-momentum-conserving to machine precision even when self-gravity is on so long as global timesteps are employed (true in all common envelope simulations published with {\sc Phantom}).

The precision of multipole methods depends mostly on the expansion order and the maximum opening angle ($p=2$ and $\theta_{\rm crit}=0.5$ in {\sc Phantom}), but also on the tree structure. Multipoles are computed on the clustering hierarchy described by its construction. To optimise the precision of a multipole expansion, the most important aspect is to compute the multipoles on the smallest volume for each of the nodes. Hence, tree construction should favour creation of compact nodes. The choice of volume splitting strategy then has a significant impact on the accuracy.

We observed this in our tests on the Plummer sphere. While the two methods used different node-node interactions, they also used different tree structures. The non-symmetric case used the previous centre-of-mass splitting for the tree construction. This gives worse maximum relative errors than splitting on the geometrical centre. The higher errors come from elongated leaf nodes produced at the edge of the Plummer core where the density gradient is strongest. Elongated nodes produce a large interaction sphere with multipoles computed on a small number of particles. Also, the local expansion of a few particles is done very far away from the centre of expansion. While the geometrical centre splitting produces deeper trees that take longer to traverse, the elimination of elongated nodes reduces the error. 

This is also confirmed by the homogeneous sphere test. There is no strong density gradient in such a sphere. Consequently, the centre of mass is always close to the geometric centre of a node, avoiding any creation of elongated boxes with the previous scheme. The resulting maximum errors are thus far below the previous distribution. This should similarly apply to our new splitting strategy; however, in this case, we observe significantly higher maximum errors. This tells us that these higher errors have a different origin. We tested our new scheme without the force correction to conserve angular momentum, and the maximum error levels dropped to a similar level to the previous scheme. In the homogeneous sphere test, the maximum errors always occur close to the centre of the distribution, where the interactions compensate for each other and leave an almost null acceleration. In such cases, the direct sum can resolve the very small acceleration. However, the FMM rapidly reaches a floor. This translates into large relative errors in these regions. Adding a force correction on these very sensitive particles significantly amplifies these larger errors, giving the result we see in \cref{fig:sphere_err} (bottom left). This effect is not visible in the Plummer case as the central region where forces compensate is significantly smaller in volume. It is less probable to generate a particle there with random generation. Moreover, in the outer edges of a Plummer, the forces also converge to zero, but their direction is still radial against the centre of the distribution. Therefore, the error amplification due to the force corrections is less important, as they will not drastically change the direction of the original forces.

The structure of our new algorithm closely follows the previous implementation in {\sc Phantom} as described in \citet{Price2018}. The symmetrisation itself thus required only small changes in the codebase. The underlying aim was to ensure similar or better performance and scaling by keeping the self-gravity calculation computed alongside the SPH force routine, rather than separately as in the \textsc{gadget4} implementation \citep{Springel2021}. This choice allows us to keep the already existing optimised particle cache that accelerates the SPH loop significantly. 

The main parallelisation strategy also remained the same, which again saved development effort. The heavier lift was to adapt this approach to recover the original $\mathcal{O}(N)$ scaling of the FMM. Our new node cache successfully achieves this goal, as we can observe in \cref{fig:plummer_perf2}. It is still important to note that this scaling is only reached when threads can work cooperatively to prune their work as the full traversal is completed. This obviously needs good scheduling from the parallelisation framework (here OpenMP). At the time, we observe the best performance with the ``dynamic'' built-in OpenMP scheduling. However, the ``static'' option can still recover the $\mathcal{O}(N)$ performance with only between $\sim 1$ and $10\%$ slower than the ``dynamic'' option. Further investigation should be done to optimise the scheduling and maximise the cache hits during the parallel traversals. \cref{fig:plummer_perf2} also shows a decrease in wall time per particle at larger $N$ with the previous scheme. This counter-intuitive behaviour is certainly a result of the efficient particle cache already present in the older version of the code, which can significantly speed up the short-range interactions at larger $N$. 

Further optimisation may accelerate other parts of the SPH loop. Currently, the density loop uses the old tree traversal to find the neighbour used to find the smoothing length of each particle. \Cref{alg:DTT,alg:cachewrite,alg:pruning} may be used in this loop to pre-cache the gravitational tensor fields and the interaction list of each node. This would result in only one ``expansive'' neighbour search, while the next ones could use the cache to prune most of the traversal. Another optimisation could use the symmetric interactions to compute and cache only one side of them (\citetalias{Dehnen2002}). This new algorithm has not been extended to the MPI domain decomposition of the code. The previous scheme is used when MPI is on. This point could be addressed if the community request it.

\section{Conclusions}\label{sec:conclusion}
We implemented a new parallel algorithm based on \citetalias{Dehnen2000} and \citetalias{Marcello2017} that symmetrises node-node interactions with adaptive gravitational softening in \textsc{Phantom}. By construction, this symmetrisation ensures that linear and angular momentum are conserved to machine precision while still using multipole expansions to approximate the self-gravity. Our tests demonstrate that this is achieved in practice. Specifically: 
\begin{enumerate}[(i)]
\item The new method gives similarly accurate gravitational forces compared to the previous default scheme in {\sc Phantom}, although this depends on the particle distribution.
\item Our new method eliminates centre of mass drift in stellar orbits and common envelope simulations while giving similar results to the previous tree-based gravity solver.
\item We modified the dual tree traversal to check and compute node-node interactions in parallel on individual branches of the tree. We also modified the acceptance criterion to account for adaptive gravitational softening without changing the Green's function in multipole expansions.
\item We avoid the need for task-based parallelism by introducing work duplication. This allowed us to implement the new traversal in the existing codebase without changing the parallelisation strategy. In top of that, we developed a caching strategy to minimise the duplication and recover the original $\mathcal{O}(N)$ scaling of the FMM.
\end{enumerate}
The overall accuracy and numerical performance of our new implementation, in addition to its machine precision momentum conservation, offer advantages over the existing method. The new method is therefore now the default solver available in \textsc{Phantom}. Individual timestepping is still breaking these conservations. However, future developments could resolve this issue by using method from \citet{Rantala_2021} for example.

\begin{acknowledgements}
We thank Walter Dehnen, Matthew Bate and Estelle Moraux for useful discussions. Computations were performed using the GRICAD infrastructure (https://gricad.univ-grenoble-alpes.fr), supported by Grenoble research communities. M. Y. M. L. acknowledges funding by the European Union (ERC, ExCEED, project number 101096243). Views and opinions expressed are, however, those of the authors only and do not necessarily reflect those of the European Union or the ERC. Neither the European Union nor the granting authority can be held responsible for them. DP also thanks IPAG, CNRS, Université-Grenoble-Alpes, François M\'enard, Nicolas Cuello and the ODYSSEY/Stellar-MADE teams for hosting and support during his sabbatical in Grenoble, where this work was completed. His visit was enabled by funding from the Campagne d’accueil de scientifiques étrangers -- UGA et Grenoble INP-UGA -- Année 2025 and by EU funding of the Stellar-MADE (project number 101042275) and DUST2PLANETS (project number 101053020) ERC projects.
\end{acknowledgements}

\bibliographystyle{aa}
\bibliography{fmm.bib}

\appendix

\section{Moment translation formulae}

\subsection{Mass Moment translation} \label{sec:massmomtrans}

The methods presented above need to compute the mass moments of nodes defined in the system domain to proceed to the multipole expansion of the Green's function. While it is possible to compute any mass moment of a node by only providing its particle set and centre. Nodes in hydrodynamical simulations are typically organised in a hierarchical tree to optimise neighbour search. Hence, nodes corresponding to internal nodes of the tree share their set of particles with their children. In such a configuration, it is numerically more efficient to compute only the mass moment on the children and translate them to the parent centre, summing the contribution of each of them.
Using this translation method, the moment of the parent node $B'$ is 
\begin{align}
    {Q_n^{B'}} & = \int_V \rho(\bm{x}_j) \bm{b'}^{(n)} ~{\rm d}^3\bm{x},\\
    & = \int_V \rho(\bm{x}_j) \left(\bm{g} + \bm{b}\right)^{(n)} ~{\rm d}^3\bm{x},\\
    & = \int_V \rho(\bm{x}_j) \underbrace{\left(\bm{g} + \bm{b}\right) \otimes \cdots  \otimes \left(\bm{g} + \bm{b}\right)}_{n \text{ times}} ~{\rm d}^3\bm{x}_j, \label{eq:massmomtrans}
\end{align}
where $\bm{g}$ is the vector between child and parent centre and $\bm{b}$ is the distance between a mass element and the centre of the child. The mass moment can then be computed only on the leaf nodes and then translated into every internal node of the tree in a bottom-up fashion.

\subsection{Gravitational field tensor translation} \label{sec:gravmomtrans}

When the multipole expansion is used to compute gravitational field tensors between pairs of nodes, the gravitational forces are passed down to every particle in a node. The simplest solution would be to proceed like that for every node in the domain. However, as mentioned above, when the symmetric fast multipole method is used, internal node-node interactions contribute to the long-range interactions of leaf nodes. In such a case, it is numerically more efficient to translate gravitational field tensors and propagate them downward through the tree, node to node, down to the leaf nodes and then to the particles.

To find the gravitational field tensor translation formula, we can express the FMM force as a function of the 3 vectors involved in the expansion
\begin{align}
    f_{\rm g}(\bm{a}_i, \bm{r}, \bm{b}_i) &=   \sum_{k = 0}^p  \frac{\bm{a}_i^{(k)}}{k!} \cdot \underbrace{(-1)^k\sum_{n=0}^{p-k} \frac{1}{n!} D_{n+k+1} (\bm{r}) \cdot {Q_n^B}(\bm{b}_i)}_{{\rm d}\bm{M}_k (\bm{r}, \bm{b}_i)}.
\end{align}
 Using a shifted centre of the box $\bm{a} = \bm{h} + \bm{a}'$, we want that 
\begin{align}
    f_{\rm g}(\bm{a}_i', \bm{r}', \bm{b}_i) &= f_{\rm g}(\bm{a}_i, \bm{r}, \bm{b}_i), \\
    &= \sum_{k = 0}^p  \frac{(\bm{a}_i' + \bm{h} )^{(k)}}{k!} \cdot {{\rm d}\bm{M}_k (\bm{r}, \bm{b}_i)}, \\
    &= \sum_{k = 0}^p   \frac{\bm{a}_i'^{(k)}}{k!} \underbrace{\sum_{l =k}^p\frac{1}{ (l-k)!}  \bm{h}^{(l-k)} \cdot {{\rm d}\bm{M}_l (\bm{r}, \bm{b}_i)} }_{{\rm d}\bm{M}_k (\bm{r}', \bm{b}_i)}.
\end{align}
We define then the translated gravitational field tensor as being 
\begin{align}\label{eq:gravmtranslat}
    {{\rm d}\bm{M}_k (\bm{r}', \bm{b}_i)} &= \sum_{l =k}^p\frac{1}{ (l-k)!}  \bm{h}^{(l-k)} \cdot {{\rm d}\bm{M}_l (\bm{r}, \bm{b}_i)},
\end{align}
under a translation of the centre by a vector $\bm{h}$. However, it is worth noting that this formula is not exact since it would require a Taylor expansion of $D_{n+k+1}$, which would require the order $2p$ of the Green's function moments. In reality, this translation operation gives the same result as the direct method up to the floating-point error.

\end{document}